\documentclass[12pt,oneside]{article}
\usepackage{amsmath,amssymb,graphics,mathtools}
\newcommand{\field}[1]{\mathbb{#1}}

\begin{document}
\title{
Justifying Typicality Measures of Boltzmannian Statistical Mechanics and Dynamical Systems}
\author{Charlotte Werndl, c.s.werndl@lse.ac.uk\\ Department of Philosophy, Logic and Scientific Method\\ London School of Economics}

\date{This article is forthcoming in: \\ \emph{Studies in History and Philosophy of Modern Physics}\\
http://www.journals.elsevier.com/studies-in-history-and-philosophy-of-science-part-b-studies-in-history-and-philosophy-of-modern-physics/}

\maketitle

\begin{abstract}
A popular view in contemporary Boltzmannian statistical mechanics is to interpret the measures as typicality measures. In measure-theoretic dynamical systems theory measures can similarly be interpreted as typicality measures. However, a justification why these measures are a good choice of typicality measures is missing, and the paper attempts to fill this gap. The paper first argues that Pitowsky's (2012) justification of typicality measures does not fit the bill. Then a first proposal of how to justify typicality measures is presented. The main premises are that typicality measures are invariant and are related to the initial probability distribution of interest (which are translation-continuous or translation-close). The conclusion are two theorems which show that the standard measures of statistical mechanics and dynamical systems are typicality measures. There may be other typicality measures, but they agree about judgements of typicality. Finally, it is proven that if systems are ergodic or epsilon-ergodic, there are uniqueness results about typicality measures.

\end{abstract}

\newpage
\tableofcontents

\newpage
\section{Introduction}
\noindent Consider all the possible states of a system in \emph{Boltzmannian statistical
mechanics}. A measure is defined over these states. The main aim of statistical mechanics is to work as a bridge between the microscopic and macroscopic levels of description of a system, and the measure is a crucial ``level-bridging'' ingredient. An important question in the foundations of Boltzmannian statistical mechanics is how to interpret this measure. A recent proposal is that it is best interpreted as a \emph{typicality measure}: it represents the \emph{relative size of sets of states}, and typical states show a certain property if the measure of the set that corresponds to this property is one or close to one. In Boltzmannian statistical mechanics the same measure is appealed to in the equilibrium and non-equilibrium context. Hence the interpretation as a typicality measure is relevant for both contexts. This approach enjoys particular prominence among contemporary physicists who endorse Boltzmannian statistical mechanics (D\"{u}rr 1998; Goldstein 2001; Goldstein and Lebowitz 2004; Lebowitz 1993a; Lebowitz 1993b; Lebowitz 1999) but has also been advocated by philosophers (Maudlin 2007; Volchan 2007).\\

\noindent Not only in statistical mechanics but also for \emph{measure-theoretic dynamical systems} an interpretation needs to be found for the standard measures used, and one suggestion is to interpret them as typicality measures. Unlike statistical mechanics, measure-theoretic dynamical systems theory is not concerned with bridging scales of description. Moreover, while statistical mechanics is a physical theory, dynamical systems theory is a set of mathematical tools with applications in various sciences (among others, in physics, chemistry, biology, meteorology and climate science). Furthermore, understanding the behaviour of ``most'' initial conditions was always a central concern in dynamical systems theory. Hence  the notion of typicality does not cause as much controversy here as in statistical mechanics.\\

\noindent What is missing is a thorough treatment of the question why these measures are a good choice of typicality measures.\footnote{Physicists sometimes seem to hold that it is fine to take the standard measures used in statistical mechanics as a postulate. They might hold such a pragmatic viewpoint because it does not lead to wrong conclusions for the practical problems they are interested in, and they do not have time to engage in more thorough foundational debates. Still, philosophers and some physicists want to arrive at a conceptual understanding of statistical mechanics, and hence are interested in a thorough conceptual motivation of typicality measures.} Witness Volchan (2007, 13): ``Many tough questions are still to be addressed.  For example, as typicality is relative to the measure used, how one justifies a particular choice? Under what criteria?''. This paper attempts to contribute to fill this gap by presenting \emph{a first tentative proposal of how to justify typicality measures in statistical mechanics and in measure-theoretic dynamical systems theory more generally}. An argument will be called a justification of typicality measures exactly when (i) the premises imply that there is a unique typicality measure or a set of typicality measures which agree on which sets are typical/atypical (here any measure of this set can be used as typicality measure); and (ii) the premises are at least potentially more plausible than just postulating that certain measures are a good choice of typicality measures.\\

\noindent This paper proceeds as follows. Section~\ref{DynamicalSystems1} introduces Boltzmannian statistical mechanics and Section~\ref{DynamicalSystems2} introduces measure-theoretic dynamical systems. Section~\ref{Typicality} discusses the notion of typicality. Section~\ref{Pitowsky} outlines Pitowsky's (2012) justification of typicality measures, which is the only paper known to the author which advances a justification of typicality measures of statistical mechanics and measure-theoretic dynamical systems. Section~\ref{PitowskyCriticism} argues that  Pitowsky's argument does not fit the bill. Section~\ref{New1} and \ref{New2} present a first proposal of how to justify typicality measures of Boltzmannian statistical mechanics and measure-theoretic dynamical systems. Finally, Section~\ref{Uniqueness} shows that if systems are ergodic/epsilon-ergodic, one obtains uniqueness results about the typicality measures. The paper concludes in Section~\ref{Conclusion}.

\section{Boltzmannian Statistical Mechanics}\label{DynamicalSystems1}
\noindent
The concern in this paper is Boltzmannian statistical mechanics\footnote{For a detailed discussion of
Boltzmannian SM see Frigg (2008) and Uffink (2007).} (SM) and not Gibbsian SM. (I adopt the common distinction that while Gibbsian SM is about ensembles, Boltzmannian SM is about a single system. So, e.g., equilibrium is associated with a single state in Boltzmannian SM and with a probability distribution in Gibbsian SM (cf.\ Frigg 2008; Uffink 1996)).\footnote{How Boltzmannian SM relates to Gibbsian SM, and why the very same measures are used in both frameworks, are deep and interesting foundational questions, which are beyond the scope of this paper (for more on his issue see Frigg 2008 and Uffink 2007).}
In Boltzmannian statistical mechanics the object of study is a system consisting of $n$ classical particles. The boundary conditions assumed here are that the system is in a bounded container and is isolated from the environment. The \emph{microscopic description} is as follows.
The \emph{microstate} of the system is represented by a point $x=(p_{1},\ldots p_{n},q_{1},\ldots q_{n})$, where $q_{i}$ is the (three-dimensional) position of the $i$-th particle and $p_{i}$ is the (three-dimensional) momentum of the $i$-th particle ($1\leq i \leq n$). The microstates $x$ are elements of the $6n$-dimensional state space $\Gamma$ (where $\Gamma$ represents all possible position and momentum coordinates of all the particles). Because the energy is conserved, the motion of the system is confined to a $(6n-1)$-dimensional energy hypersurface $\Gamma_{E}$, which can be shown to be compact ($E$ is the value of the energy of the system). The system starts in a certain initial condition -- an initial microstate $x$. The evolution of the system is governed by Hamilton's equations, whose solutions are the phase flow $\phi_{t}$ on the energy hypersurface $\Gamma_{E}$. That is, $\phi_{t}(x)$ gives the microstate of the system that started in initial condition $x$ after $t$ time steps. \\

\noindent The \emph{macroscopic description} is as follows. \emph{Macrostates} $M_{i}$, $1\leq i\leq n$, are characterised by the values of macroscopic parameters such as local pressure and local temperature. To each macrostate $M_{i}$, $i=1,\ldots, k$ $(k\in\field{N})$, there corresponds a \emph{macroregion} $\Gamma_{M_{i}}$ which consists of all $x\in \Gamma_{E}$ for which the macroscopic parameters take values which are characteristic for $M_{i}$. The $\Gamma_{M_{i}}$, $1\leq i \leq k$, do not overlap and jointly cover $\Gamma_{E}$.\footnote{That is, technically the $\Gamma_{M_{i}}$ are a \emph{partition}, meaning that $\Gamma_{M_{i}}\in\Sigma_{\Gamma_{E}}$ for all $i$, $\Gamma_{M_{i}}\cap \Gamma_{M_{j}}=\emptyset$ for $i\neq j$ and $\cup_{i=1}^{k}\Gamma_{M_{i}}=\Gamma_{E}$.} Two macrostates are of particular importance: the equilibrium state $M_{eq}$ and the macrostate at the beginning of the process $M_{p}$. The macroscopic evolution of the system with initial microstate $x$ is given by $M_{x}(t)=M(\phi_{t}(x))$, where $M(y)$ is the macrostate corresponding to microstate $y$.\\

\noindent Regarding the measures of interest in Boltzmannian SM, note that for Hamiltonian systems the uniform measure $\mu$ defined on $\Gamma$ (the Lebesgue measure) is invariant under the dynamics; this result is known as Liouville's theorem (Petersen 1983, 5). $\mu$ can be restricted to a normalized measure $\mu_{E}$ on $\Gamma_{E}$, which is also invariant under the dynamics. $\mu_{E}$ is called the \emph{microcanonical measure} and is the standard measure used in Boltzmannian SM. Hence the question is how to justify $\mu_{E}$ as typicality measure. \\

\noindent Because the same measures are employed in Bolzmannian SM for questions regarding equilibrium and non-equilibrium, the interpretation as typicality measure is relevant in both contexts.\footnote{For instance, it is often argued that typical initial states show thermodynamic-like behaviour (a claim about non-equilibrium) or that typical states are in equilibrium (a claim about equilibrium).} In general, much more is known about equilibrium SM than non-equilibrium SM.\footnote{In practice physicists usually employ Gibbsian SM, and here the standard measures (microcanonical measures, canonical measures, and grand-canonical measures) are very successful in deriving predictions about macroscopic behaviour. The relationship between Boltzmannian and Gibbsian SM is a controversial theme, which goes beyond the scope of this paper (for more see Frigg 2008 and Uffink 2007).}
In particular, it is a central aim of non-equilibrium SM to explain \emph{thermodynamic-like behaviour}, but this is extremely difficult. To characterise thermodynamic-like behaviour, the Boltzmann entropy first needs to be introduced. The \emph{Boltzmann entropy of a macrostate} $M_{i}$ is $S_{B}(M_{i})= k_B \log(\mu_{E}(\Gamma_{M_{i}}))$, where $k_B$ is the Boltzmann constant; and the \emph{Boltzmann entropy of a system} at time $t$ is $S_{B}(t)= S_{B}(M_{x(t)})$.
It can be shown that for gases the equilibrium macroregion covers nearly all of $\Gamma_{E}$, and hence the Boltzmann entropy is highest in equilibrium.
The macrostate at the beginning of the process $M_{p}$ is, by assumption, a low entropy state. Thermodynamic-like behaviour is characterised as follows (Lavis 2005): the general tendency is that the entropy of a system that is initially in a low-entropy state increases until it comes close to its maximum value and then stays there, but frequent small and very rare large downward fluctuations (contra irreversibility) are allowed.  Now proponents of the typicality approach argue that typical initial states show thermodynamic-like behaviour, and hence that \emph{an explanation of thermodynamic-like behaviour can be given in terms of typicality}.\footnote{For a list of proponents of the typicality approach, see the references given in the introduction.}

\section{Measure-Theoretic Dynamical Systems}\label{DynamicalSystems2}

\noindent Unlike statistical mechanics, measure-theoretic dynamical systems theory is not concerned with bridging scales of description. Moreover, while statistical mechanics is a physical theory, dynamical systems theory is a set of mathematical tools with applications in various sciences (among others, in physics, chemistry, biology, meteorology and climate science). Furthermore, understanding the behaviour of ``most'' initial conditions was always a central concern in dynamical systems theory. Hence  the notion of typicality does not cause as much controversy here as in statistical mechanics.\\

\noindent
Dynamical systems theory can be split into measure-theoretic dynamical systems theory (where there is a measure and statistical properties are studied) and topological dynamical systems theory (where there is no measure and topological properties are studied).\footnote{Dynamical systems have been studied from both the measure-theoretic and the topological perspective, and the interrelations between these perspectives can be very complex (cf.\ Petersen 1983).} \emph{This paper is only concerned with measure-theoretic dynamical systems theory}. There are also interesting foundational questions in topological dynamical systems theory, where a topological notion of typicality plays a role, but this topic is beyond the scope of this paper (see, e.g., Frigg and Werndl, 2012, for some thoughts on topological typicality).\\

\noindent
Formally, measure-theoretic dynamical systems (mt-dynamical systems) are defined as follows (Petersen 1983). $(X,\Sigma_{X},\mu_{X},f_{t})$ is a \emph{mt-dynamical system} iff (if and only if) $X$ is a set (the \emph{phase space}), where in this paper $X$ is an interval\footnote{An interval in $\field{R}^{n}$ is any set $A_{1}\times \ldots \times A_{n}$ with $A_{i}=(a_{i},b_{i})$ ($a_{i},b_{i}\in\field{R}\cup\{-\infty,\infty\}$), $(a_{i},b_{i}]$ ($a_{i}\in\field{R}\cup\{-\infty,\infty\}$, $b_{i}\in\field{R}$), $[a_{i},b_{i})$ ($a_{i}\in\field{R}$, $b_{i}\in\field{R}\cup\{-\infty,\infty\}$), or $[a_{i},b_{i}]$ ($a_{i},b_{i}\in\field{R}$), where always $a_{i}<b_{i}$, $1\leq i \leq n$.} in $\field{R}^{n}$; $\Sigma_{X}$ is the Lebesgue $\sigma$-algebra of subsets of $X$ (elements of $\Sigma_{X}$ are called \emph{measurable} sets); $\mu_{X}:\Sigma_{X}\rightarrow [0,1] $ is a normalized measure on $X$; and $f_{t}:X\rightarrow X$ (the \emph{evolution equations}, $f_{t}(x)$ denotes the state of the system that started in initial state $x$ after $t$ time steps), where $t\in\field{R}^{+}_{0}$ (continuous time) or $t\in\field{N}_{0}$ (discrete time), are surjective measurable mappings such that $f_{t+s}=f_{t}(f_{s})$ for all
$t,s\in\field{R}^{+}_{0}$ or $\field{N}_{0}$, $f_{t}(x)$ is jointly measurable in $(x,t)$, and the measure is \emph{invariant under the dynamics}, i.e.,
\begin{equation}\label{invariance}
\mu_{X}(A)=\mu_{X}(f_{t}^{-1}(A))\,\,\textnormal{for all measurable}\,\,A\subseteq X\,\,\textnormal{and all}\,\,t\in\field{R}^{+}_{0}\,\,\textnormal{or}\,\, \field{N}_{0}.\end{equation}
Statistical mechanical systems $(\Gamma_{E}, \Sigma_{\Gamma_{E}}, \mu_{E},\phi_{t})$ are
(continuous-time) mt-dynamical systems (where $\Sigma_{\Gamma_{E}}$ is the Lebesgue $\sigma$-algebra of $\Gamma_{E}$). This just means that the $(\Gamma_{E}, \Sigma_{\Gamma_{E}}, \mu_{E},\phi_{t})$ satisfy the formal definition of a mt-dynamical system. It does \emph{not} imply that SM is reducible to dynamical systems theory in any sense (the physical postulates of SM are not part of the mathematical framework of dynamical systems theory). \\

\noindent Let me introduce two (discrete-time) mt-dynamical systems which will serves as examples. They are very simple and hence particularly suited for illustration purposes.\\

\noindent
\noindent\emph{Example 1.} The first example is the \emph{tent
map} $(Y,\Sigma_{Y},\mu_{Y},s_{t})$ where $Y=[0,1]$ is the unit interval, $s_{t}(x)=s^{t}(x)$ (i.e., the $t$-th iterate\footnote{The $0$-th iterate is the identity function, i.e., $s^{0}(x)=x$ for all $x\in X$.} of $s(x)$), $t\in\field{N}_{0}$, with
\begin{equation}
s(x)=
\begin{cases} 2x & \text{if $0\leq
x<\frac{1}{2}$,}
\\
2(1-x) &\text{if $\frac{1}{2}\leq
x\leq 1$,}
\end{cases}
\end{equation}
and $\mu_{Y}$ is the uniform measure (Lebesgue measure). Figure 1(a) shows the density of the measure $\mu_{Y}$ of the tent map. \\

\begin{figure}
\centering
\includegraphics{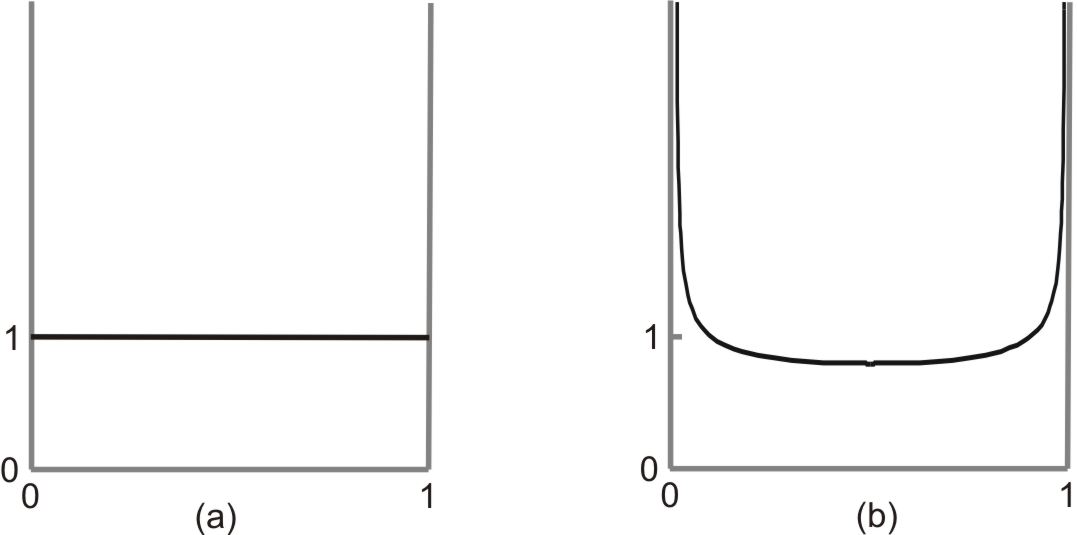}
\caption{(a)$\,\,$the$\,\,$measure$\,\,$of$\,\,$the$\,\,$tent$\,\,$map;$\,\,$(b)$\,\,$the$\,\,$measure$\,\,$of$\,\,$the$\,\,$logistic$\,\,$map}\label{billiard}
\end{figure}

\noindent \emph{Example 2.} The second example is the \emph{logistic map}
$(Z,\Sigma_{Z},\mu_{Z},r_{t})$ where $Z=[0,1]$, $r_{t}(x)=r^{t}(x)$,
$t\in\field{N}_{0}$, with
\begin{equation}
r(x)=4x(1-x),
\end{equation} and the standard measure is given by
\begin{equation}
\mu_{Z}(A)=\int_{A}\frac{1}{\pi\sqrt{x(1-x)}}d\lambda,
\end{equation}
where $\lambda$ is the Lebesgue measure. Figure 1(b) shows the density of the measure $\mu_{Z}$ of the logistic map. 
The measure of the logistic map is particularly interesting because, unlike that of the tent map, it is \emph{not} uniform. \\

\noindent
The definition of mt-dynamical systems immediately implies that all dynamical systems are \emph{forward deterministic}: the state of the system at one time determines the state of the system at all future times. If, additionally, the state of the system at one time determines the state of the system at all past times, then the system is \emph{deterministic} (i.e., \emph{forward and backward deterministic}). Some mt-dynamical systems such as Example 1 (the tent map) or Example 2 (the logistic map) are only forward deterministic. Other mt-dynamical systems such as those in SM are deterministic.

\section{Typicality}\label{Typicality}
\noindent
A typicality measure represents the \emph{relative size of sets of states}. Intuitively speaking, any function $\mu_{T}$ describing the size of sets of states should satisfy the standard axioms of a \emph{measure}: $\mu_{T}(\emptyset)=0$, $\mu_{T}(A)\geq 0$ for any measurable set $A$, $\mu_{T}(A\cup B)=\mu_{T}(A)+\mu_{T}(B)$ whenever $A\cap B=\emptyset$ where $A, B$ are measurable sets, and \label{sad} $\mu_{T}(\cup_{i=1}^{\infty}A_{i})=\sum_{i=1}^{\infty}\mu_{T}(A_{i})$ whenever $A_{i}\cap A_{j}=\emptyset$ for all $i,j,i\neq j$, where all $A_{i}$ are measurable sets. One might also argue that, because typicality measures represent the \emph{relative} size of sets of states, they should be normalized, i.e., $\mu_{T}(C)=1$, where $C$ is the set of all elements. Normalized measures are called \emph{probability measures}. In what follows, I assume that typicality measures are normalized. However, this assumption is irrelevant for the main result in this paper (Theorems 1 and 2), where there is exactly the same conclusion if the measures are not normalized.\footnote{It is easy to see that the proof of Theorem 1 and 2 also goes through when the measures are not normalized.}  The only results which depend on the assumption that typicality measures are normalized are the uniqueness results (Theorem 3 and 4), which invoke the notion of ergodicity and epsilon-ergodicity.\\

\noindent
Intuitively speaking, something is typical if it happens in the vast majority of cases. For instance, typically, when throwing dice, at some point the number six will show up;
or typical lottery tickets are blanks.
Formally, given a basic set of elements $C$, one says that \emph{typical elements have property P} relative to a typicality measure $\mu_{T}$ iff \emph{$\mu_{T}(C\setminus D)=0$}, where $D\subseteq C$ consists of those elements in $C$ that have property $P$ (this will be called \emph{typicality I}). Derivatively, $D$ is called the \emph{typical set} and $C\setminus D$ the \emph{atypical set}. A less stringent notion of typicality arises when one only requires that the relative size of atypical sets is close to zero (and not zero). Formally, given a basic set of elements $C$, one then says that \emph{typical elements have property P} relative to a typicality measure $\mu_{T}$ iff \emph{$\mu_{T}(C\setminus D)\leq \beta$}, where $D\subseteq C$ consists of those elements in $C$ that have property $P$ and $\beta$ is a very small real number (this will be called \emph{typicality II}). Then, again, $D$ is called the \emph{typical set} and $C\setminus D$ the \emph{atypical set}.\\

\noindent
For example, in SM it is claimed that $\mu_{E}$ is the typicality measure and that typical initial states show thermodynamic-like behaviour. That is, $C=\Gamma_{E}, \mu_{T}=\mu_E$, $P$ is the property of showing thermodynamic-like behaviour and $D$ is the set of initial states showing thermodynamic-like behaviour. Measures of mt-dynamical systems, such as the uniform measure $\mu_{Y}$ of Example 1 of the tent map (see Figure 1(a)) and the measure $\mu_Z$ of Example 2 of the logistic map (see Figure 1(b)), are also candidates for typicality measures.\\

\noindent
Part and parcel of the notion of typicality measures is that states cannot become more or less when they are evolved under the dynamics of the system. Formally, this means that they should be \emph{invariant under the dynamics: the typicality measure of a set has to equal (i.e., cannot be greater or smaller than) the typicality measure of the evolved set.} Formally, for evolution equations such as those in SM, which are invertible (and hence forward and backward deterministic), this amounts to the requirement that the size of a set of states $A$ should be the same as the size of the image or pre-image of $A$:
\begin{equation}\label{I2}
\mu_{T}(A)\!=\!\mu_{T}(f_{t}(A))\!=\!\mu_{T}(f_{t}^{-1}(A))\,\,\textnormal{for all}\,\,t\!\in\! \field{R}^{+}_{0}\,\,\textnormal{or}\,\,t\in\!\field{N}_{0}\,\,\textnormal{for all measurable}\,\,A.
\end{equation}
For evolution equations such as those of Example 1 (the tent map) or Example 2 (the logistic map), which are not invertible (hence only forward deterministic), the requirement is that the size of a set of states $A$ should be the same as the size of the pre-image of $A$:
\begin{equation}\label{I1}
\mu_{T}(A)=\mu_{T}(f_{t}^{-1}(A))\,\,\textnormal{for all}\,\,t\in \field{R}^{+}_{0}\,\,\textnormal{or}\,\,t\in\field{N}_{0}\,\,\textnormal{for all measurable}\,\,A.
\end{equation}
Proponents of the typicality approach also require this (e.g., Goldstein 2001, 15;
Lebowitz 1993b, 8).\footnote{This invariance condition is so important in measure-theoretic dynamical systems theory that for a mt-dynamical system it is required that the measure is invariant (see equation~(\ref{invariance})).}\\

\noindent
When discussing the invariance condition, it is important to mention \emph{Liouville's equation}, which is a central equation in SM that describes the time-evolution of a probability density  $\rho(x)=\rho(p_{1},\ldots,p_{n},q_{1},\ldots,q_{n})$ in phase space:
\begin{equation}
\frac{\partial \rho}{\partial t}=-\sum_{i=1}^{n}\frac{\partial \rho}{\partial q_{i}}\dot q_{i}+\frac{\partial\rho}{\partial p_{i}}\dot p_{i},
\end{equation}
where the time-derivatives are denoted by dots. The notion of absolute continuity will be important in what follows. Formally, for measures $\mu_1$ and $\mu_2$ defined on $(X,\Sigma_{X})$, $\mu_2$ is \emph{absolutely continuous w.r.t.\ $\mu_1$} $(\mu_2\ll\mu_1)$ iff
\begin{equation}\label{ac}
\textnormal{if}\,\, \mu_{1}(A)=0\,\,\textnormal{for a measurable set}\,\,A,\,\,\textnormal{then}\,\,\mu_{2}(A)=0.
\end{equation}
The stationary solutions of the Liouville equation (i.e., when $\partial \rho/\partial t=0$) give the probability densities which do not change in time and hence the invariant measures absolutely continuous w.r.t.\ the Lebesgue measure.\footnote{According to the Radon-Nikodym theorem, a measure is absolutely continuous w.r.t.\ the Lebesgue measure iff there is a probability density (Nielsen, 1997).} Finding the stationary solutions of the Liouville equation for systems in SM is extremely hard. In general, one does not know the class of all its stationary solutions. In Section 7 where a justification of typicality measures for typicality I is proposed, it is argued that typicality measures are invariant and absolutely continuous w.r.t.\ the Lebesgue measure.\footnote{The latter claim follows from Premises 2 and 3.} Therefore, the stationary solutions of the Liouville equation are of interest because they are potential candidates for typicality measures.\\

\noindent
The next section will outline Pitowsky's (2012) argument, which is the only justification of typicality measures of SM and measure-theoretic dynamical systems known to the author.

\section{Pitowsky's Justification of Typicality Measures}\label{Pitowsky}
\noindent
Pitowsky's argument goes as follows. Consider $\{0,1\}^{n}$ -- the set of all sequences of length $n$ consisting of zeros and ones ($n\in\field{N}$).  When the phase space consists of a \emph{finite} number of $N$ elements ($N\in\field{N}$), Pitowsky claims that the typicality measure is easy to find: just count the number of elements of a set and divide it by $N$. Hence, for an arbitrary subset $A$ of $\{0,1\}^{n}$, the typicality measure is $\mu_{n}(A)=|A|/2^{n}$, where $|A|$ is the number of elements of $A$. Pitowsky states that the real difficulty lies in the question of how to determine typicality measures for phase spaces with an \emph{infinite} number of elements (as in SM and measure-theoretic dynamical systems theory). Pitowsky does not provide any arguments why the uniform measure is the correct typicality measure for phase spaces with a finite number of elements. \emph{It is important to point out that this claim can be questioned}. I will argue that even if this claim is accepted, Pitowsky's argument fails.\\

\noindent
To tackle the question how to justify typicality measures for phase spaces with an infinite number of elements, Pitowsky starts with the special case of $\{0,1\}^{\infty}$ -- the set of all infinite sequences of zeros and ones. A unique measure $\mu_{\infty}$ on $\{0,1\}^{\infty}$ is
obtained by approximating a set $B$ in $\{0,1\}^{\infty}$ by sets $A$ in $\{0,1\}^{n}$,
$n\in\field{N}$, and then defining the measure of $B$ to be the limit of the measure of
the approximating sets $A$. The measure $\mu_{\infty}$, Pitowsky claims, it the
correct typicality measure on $\{0,1\}^{\infty}$. \\

\noindent
How to arrive at typicality measures on phase spaces of SM and measure-theoretic dynamical
systems theory, which are usually quite different from $\{0,1\}^{\infty}$?
Pitowsky (2012) concentrates on the case where \emph{the phase space is the unit interval $[0,1]$}. He claims that all other cases can be reduced to this case because any relevant\footnote{Technically, the condition is that the measure space is a Lebesgue space (for a definition see Petersen 1983).} phase space with a measure on it is isomorphic to the measure space of the unit interval $[0,1]$ with the uniform measure on it. To spell out my criticism of Pitowsky (2012), only a
discussion of this special case of the unit interval will be needed. Thus, for simplicity, what follows concentrate on it. (However, I will briefly comment on Pitowsky's appeal to the isomorphism result in footnote~\ref{isom}).\\

\noindent
To arrive at a typicality measure on $[0,1]$, Pitowsky first notes that
the points of $\{0,1\}^{\infty}$ can be put into one-to-one correspondence with the points
of $[0,1]$ by assigning to each sequence $\omega=(\omega_{1},\omega_{2},\ldots)$,
$\omega\in \{0,1\}^{\infty}$, the number\footnote{Strictly speaking, the function only establishes a one-to-one correspondence of $\{0,1\}^{\infty}\setminus Q$ and $[0,1]$, where $Q$ is the
set of all sequences ending with an infinite number of ones excluding the sequence $(1,1,\ldots)$ consisting only of ones. That $Q$ is excluded
is irrelevant because $\mu_{\infty}(Q)=0$ (cf.\ Pitowsky 2012).}
\begin{equation}
\phi(\omega)=\sum_{i=1}^{\infty}\omega_{i}/2^{i}.
\end{equation} $\phi$ maps the measure $\mu_{\infty}$ on $\{0,1\}^{\infty}$ to a
measure on $[0,1]$. Namely, one obtains the uniform measure on $[0,1]$. Consequently, Pitowsky concludes, the uniform
measure is the uniquely correct typicality measure on $[0,1]$.

\section{Criticism of Pitowsky's Justification}\label{PitowskyCriticism}
\noindent
Unfortunately, problems arise with Pitowsky's justification. First of all, Pitowsky arrives at a measure on $[0,1]$ by using $\phi$ to map the measure $\mu_{\infty}$ on
$\{0,1\}^{\infty}$ to a measure on $[0,1]$. However, \emph{other functions than $\phi$ which provide a one-to-one correspondence between the points of $\{0,1\}^{\infty}$ and $[0,1]$ could be used to map the measure on $\{0,1\}^{\infty}$ to a measure on $[0,1]$.} Indeed, one can arrive at infinitely many different measures on $[0,1]$ by relating $[0,1]$ to the set of sequences of zeros and ones.
For example, suppose that to each sequence
$\omega=(\omega_{1},\omega_{2},\ldots)$, $\omega\in \{0,1\}^{\infty}$,
is assigned the number\footnotemark[\value{footnote}]
\begin{equation}\label{logisticC}
\psi(\omega)=\sin^{2}(\frac{\pi}{2}\sum_{i=1}^{\infty}\omega_{i}/2^{i}).
\end{equation}
Then, when $\psi$ is used to map the measure $\mu_{\infty}$ to a measure on
$[0,1]$, one obtains the invariant measure of the logistic map 
shown in Figure 1(b) 
(Aligood et al.\ 1996). \emph{To sum up, it is unclear which map should be used to arrive at a typicality measure on
$[0,1]$.} Pitowsky does not address why $\phi$ should be preferred over other maps,
and there seems to be no easy answer to the question which map to prefer. Thus Pitowsky's argument is wanting.\\

\noindent
Second, Pitowsky (2012, 17-18) explicitly states that his argument not only applies to statistical mechanical systems but more generally to many other mt-dynamical systems.
However, as I will show now, here his argument leads to undesirable conclusions. According to Pitowsky, his argument applies to the logistic map (Example 2)\footnote{Pitowsky (2012) states that his argument applies to two-symbol Bernoulli systems, and the logistic map is a two-symbol Bernoulli system (cf.\ footnote~\ref{BernoulliSystem}).}, which is defined on the unit interval. Hence for the logistic map the uniform measure is the correct typicality measure. However, \emph{the measure which dynamical system theorists' choose (i.e., the correct measure according to practice in dynamical systems theory) is different}, namely $\mu_{Z}$ (as shown in Figure 1(b)). 
Furthermore, recall that, as argued in Section~\ref{Typicality}, typicality measures should be invariant under the dynamic. However, for the logistic map the uniform measure is \emph{not} invariant and hence cannot be a typicality measure. Instead, the standard measure $\mu_{Z}$ \emph{is} invariant. (I will argue in Sections~\ref{New1}, ~\ref{New2} and 9 that $\mu_{Z}$ is the correct typicality measure).\\

\noindent
The general picture is that for some mt-dynamical systems with phase space $[0,1]$ the uniform measure is the correct measure according to practice in dynamical systems theory and is invariant under the dynamics. The tent map (Example 1) is a case in point. \emph{However, for many other mt-dynamical systems with phase space $[0,1]$, such as the logistic map (Example 2), the correct measure according to practice in dynamical systems theory is not the uniform measure and the uniform measure is not invariant}. Yet Pitowsky's argument implies that the uniform measure is always the correct typicality measure. Hence his argument is flawed. This point suggests that when justifying typicality measures not only the phase space (as in Pitowsky's justification) but also the \emph{dynamics} needs to be taken into consideration: as Example 1 (the tent map) and Example 2 (the logistic map) illustrate, different dynamics on $[0,1]$ will lead to different typicality measures. \\

\noindent
Indeed, Pitowsky (2012, 17-18) gives examples of mt-dynamical systems which highlight these problems. Intuitively speaking, \emph{two-symbol Bernoulli systems} are dynamical systems whose solutions can be put into one-to-one correspondence with the set of all infinite sequences arising from independent trials of tossing a coin (i.e., the set $\{0,1\}^{\infty}$).\footnote{
Formally, let $\Omega=\{0,1\}^{\infty}$. \label{BernoulliSystem} That is, $\Omega$ is the set of all infinite sequences $(\omega_{1}, \omega_{2},\ldots)$ with $\omega_{i} \in \{0,1\}$ corresponding to the possible outcomes of an infinite sequence of independent trials of a (possibly biased) coin. Let $\Sigma_{\Omega}$ be the set of all subsets of infinite sequences to which probabilities can be assigned, let $\mu_{\Omega}$ be the normalized measure on $\Omega$ arising from the probability measure of tossing the coin, and let $S:\Omega \rightarrow \Omega$, $S((\omega_{1}, \omega_{2}, \ldots))=(\omega_{2}, \omega_{3}, \ldots)$. $(\Omega,\Sigma_{\Omega},\mu_{\Omega},S)$ is called a \emph{two-symbol Bernoulli shift}. Finally, \emph{two-symbol Bernoulli systems} are defined to be mt-dynamical systems which are isomorphic (from a measure-theoretic perspective) to two-symbol Bernoulli shifts (cf.\ Werndl 2009a).}
Because the solutions of two-symbol Bernoulli systems correspond to the set of all infinite sequences $\{0,1\}^{\infty}$, Pitowsky argues that \emph{two-symbol Bernoulli systems and their measures illustrate the correctness of his justification.} However, contrary to Pitowsky, the measure of many two-symbol Bernoulli systems with phase space $[0,1]$ is \emph{not} the uniform measure. For instance, the logistic map (Example 2) is a two-symbol Bernoulli system and its solutions can be put into one-to-one correspondence with $\{0,1\}^{\infty}$ (see equation (\ref{logisticC})) (Werndl 2009a). Yet $\mu_{Z}$ -- the measure of the logistic map -- is not uniform. Consequently, \emph{rather than illustrating the correctness of Pitowsky's argument, two-symbol Bernoulli systems highlight its problems}.\footnote{As mentioned above, Pitowsky (2012) claims that cases where the phase space is not $[0,1]$ can be reduced to the case where the phase space is $[0,1]$ because any relevant phase space with a measure on it (technically: any Lebesgue space) is isomorphic to the measure space of $[0,1]$ with the uniform measure on it.
I do not think that Pitowskys' appeal to this isomorphism result is successful. Pitowsky's concern is to find, given a phase space $S$ such as $\Gamma_{E}$, the correct typicality measure. But then \emph{this isomorphism result is of no help because it presupposes that there is already a measure defined on S}. If just the phase space $S$ is given (this is the problem of concern), then there are usually infinitely many ways of mapping $S$ to the unit interval. Consequently, even if one assumes that the correct typicality measure on $[0,1]$ is the uniform measure, there are infinitely many different possible measures on $S$ arising from mapping the uniform measure on $[0,1]$ to $S$ (and hence infinitely many different possible measures on $S$ such that the resulting measure space is isomorphic to $[0,1]$ with the uniform measure on it.) Hence \emph{there is no hope that the isomorphism result can justify that a certain measure such as $\mu_{E}$ on $\Gamma_{E}$ is the correct typicality measure.}
\label{isom}}\\

\noindent
To conclude, Pitowsky's justification of typicality measures does not fit the bill.
The next two sections aim to provide a first tentative proposal of how to justify typicality measures in SM and measure-theoretic dynamical systems theory. Section~\ref{New1} presents the argument for typicality I and Section~\ref{New2} the argument for typicality II (see Section~\ref{Typicality} for a definition of typicality I and typicality II). The guiding idea is that typicality measures are invariant and are related to the initial probability distributions of interest (in particular, typicality measures should serve as a shortcut to make claims about any initial probability distribution of interest).

\section{A New Proposal: Typicality I}\label{New1}

First of all, as argued in Section 4, typicality measures are invariant:\vspace{2mm} \\
\noindent\textbf{Premise 1: Typicality measures are invariant under the dynamics.}\vspace{2mm} \\
\noindent It is important to note that \emph{the requirement of invariance alone cannot justify typicality measures.} To stress the point, consider the phase space $W=[0,1]$ with the evolution equation $f_{t}(w)=w$ for all $t\in\field{N}_{0}$. Clearly, for this phase space and evolution equation \emph{all} measures fulfill equation (\ref{I2}). Thus they are all invariant under the dynamics, but they do not agree on which sets are typical/atypical. Hence the condition of invariance alone cannot justify typicality measures.
\\

\noindent An \emph{initial probability distribution} gives the probability that a system is prepared in a certain microstate (e.g., that an experimenter prepares a gas in a certain microstate). This paper will assume that these initial probability distributions $p$ are \emph{translation-continuous}, i.e., that
\begin{equation}
\lim_{\left|\left|\tau\right|\right|\rightarrow 0}p(Tr(A,\tau))=p(A)\,\,\textnormal{for all open sets}\,\,A,
\end{equation}
where $\left|\left|\tau\right|\right|$ is the standard Euclidean norm in $\field{R}^{n}$ (i.e., the standard Euclidean metric between the vector $\tau$ and the zero-vector in $\field{R}^{n}$) and $Tr(A,\tau)$ is the set $A$ translated by $\tau$, i.e.,
\begin{equation}
Tr(A,\tau)=X\cap\{x+\tau\,\,|\,\,x\in A\}.
\end{equation}
Malament and Zabell (1980) have provided a nice motivation of this condition: they have argued that when one prepares a system in SM, one does not have sufficient accuracy to create any other probability measure than a translation-continuous probability measure.\footnote{
Leeds (1989) has reconstructed Malament and Zabell's argument for translation-continuity as the claim that the microstate of a system in SM is a continuous function of the parameters of the preparation of the system. As van Lith (2001) and Vranas (1998) have pointed out, this claim is problematic: clearly, the microstate of a system not only depends on the parameters of preparation but also on the microstate prior to preparation. Hence similar values of the preparation parameters may well lead to quite different microstates. However, Malament and Zabell seemed to have the different claim in mind that we do not have sufficient accuracy to create probability measures that are not translation-continuous.} Malament and Zabell (1980) have also shown that a probability measure $p$ is translation-continuous iff $p\ll\lambda$, i.e., $p$ is absolutely continuous w.r.t.\ the Lebesgue measure $\lambda$ (see equation (\ref{ac})).\footnote{They proved this result only for the phase space $\field{R}^{n}$. Yet it is clear that it also holds for any arbitrary interval of $\field{R}^{n}$. Although the phase space of systems in SM are not intervals, Malament and Zabell (1980) and Vranas (1998) believe that the equivalence of translation-continuity and absolute continuity also holds for the phase spaces in SM, and I follow them in assuming this.} That $p\ll \lambda$ seems to be often endorsed in the physics community (cf.\ Leeds 1989; Maudlin 2007).\\

\noindent
Now, different initial probability distributions might arise in SM: Systems can be prepared in different macrostates, different scientists might prepare systems differently. Thus there is a \emph{class $P$ of initial probability distributions of interest}.
Open sets are extended regions of phase space. This motivates the only additional requirement on this class, namely that for any arbitrary open region of phase space one cannot exclude that there might be some way of preparing the system such that there is a positive probability that one ends up in this region. So:\vspace{2mm}\\
\noindent \textbf{Premise 2. The initial probability distributions of interest are a class $P$ of translation-close probability distributions where for every open set $A$ there is a $p\in P$ with $p(A)>0$.}\vspace{2mm} \\
Davey (2008), Hemmo and Shenker (2012) and Leeds (1989) have criticised the extant foundational literature on Boltzmannian SM for assuming that the initial probability distribution is $\mu_{E}$ and has to be invariant under the dynamics. They rightly argue that the initial probability distribution can be different from $\mu_{E}$ and may well not be invariant under the dynamics. For instance, Hemmo and Shenker (2012, 11) complain: ``In particular the measure need not be the Lebesgue measure, and may not even be conserved under the dynamics''.\footnote{Note that this remark about the initial probability distribution could not also be made about the typicality measure. The typicality measure is not the initial probability distribution over the states in an experiment. The typicality measure counts states and, as argued, it is part and parcel of a measure that counts states that it is invariant (because states cannot become more or less when they are evolved under the dynamics).} So \emph{the great flexibility on the initial probability distribution in SM allowed by Premise 2 should be welcome}.\\

\noindent Concerning dynamical systems theory, note that in certain applications initial probability distributions over the states play a role. The justification of typicality measures advanced here will be relevant whenever Premise 2 holds. For instance, consider the dynamical system of a frictionless \emph{pendulum}, consisting of a mass $m$ that experiences a single force $F$ which depends only on the position of the mass and a constant (this is the harmonic oscillator). Suppose that one does not have sufficient accuracy to create any other initial probability distribution than a translation-continuous one (which has some plausibility). Further, since in experiments one has the freedom to prepare the system in all kinds of different initial positions and initial velocities, it seems reasonable that for any arbitrary open region of phase space there is a probability distribution of interest which assigns positive probability to this region. Then the uniform measure of the pendulum can be justified as typicality measure.\\

\noindent
The next two premises are motivated by the idea that \emph{typicality measures should be related to the initial probability distributions of interest}. More specifically, first, it seems reasonable to require that if a set of states has probability zero for all initial probabilities of interest, the set is atypical: \vspace{2mm} \\
\noindent \textbf{Premise 3. If p(A)=0 for all probability distributions of interest for some measurable A, then T(A)=0.}\\
(Equivalent formulation: If $p(A)=1$ for all probability distributions of interest for some measurable $A$, then $T(A)=1$).\\

\noindent Second, the typicality measure should serve as a shortcut to make claims about the likely behaviour of the system for any possible initial probability distribution. That is, it will be required that whenever a set of states is typical, it has probability one for all initial probability distributions of interest: \vspace{2mm} \\
\noindent \textbf{Premise 4. If $T(A)=1$ for some measurable $A$, then $p(A)=1$ for all probability distributions of interest.} \\
(Equivalent formulation: If $T(A)=0$ for some measurable $A$, then $p(A)=0$ for all probability distributions of interest).\\

\noindent
Also, the following premise is adopted:\vspace{2mm} \\
\textbf{Premise 5. Whenever a measure fulfills Premises 1, 3 and 4 (for the probability distributions as characterised by Premise 2), then it is a typicality measure}.\vspace{2mm} \\
It will be shown\footnote{This is shown when proving Theorem 1 (see Appendix 1).} that measures satisfying Premises 1, 3 and 4 agree on which sets are typical or there is a unique measure satisfying these premises. Hence it seems reasonable to take these conditions as sufficient for typicality measures.\\

\noindent Finally, let me specify the systems under consideration:\vspace{2mm} \\
\textbf{Premise 6. Let a statistical mechanical system $(\Gamma_{E},\Sigma_{\Gamma_{E}},\phi_{t},\mu_{E})$ be given, or let a dynamical system $(X,\Sigma_{X},\phi_{t},\mu_{X})$ be given where $X$ is an interval in $\field{R}^{n}$, $\mu_{X}\ll \lambda$ and $\lambda\ll \mu_{X}$.}\vspace{2mm} \\
For many standardly used measures $\mu_{X}$ in measure-theoretic dynamical systems theory $\mu_{X}\ll \lambda$ and $\lambda\ll \mu_{X}$. For instance, it is clear that this condition holds for the measure $\mu_{Y}$ of the tent map (Example 1) and the measure $\mu_{Z}$ of the logistic map (Example 2).\\

\noindent
The following result is derivable from these premises (for the proof see Appendix 1):\vspace{2mm} \\
\textbf{
Conclusion (Theorem 1). The measure $\mu_{E}$ of systems in statistical mechanics/the measure $\mu_{X}$ of dynamical systems is a typicality measure. Furthermore, any other typicality measure will agree with $\mu_{E}$/$\mu_{X}$ on which sets are typical.} \vspace{2mm} \\
This result aims to justify typicality measures $\mu_{E}$ in SM and typicality measures $\mu_{X}$ of dynamical systems such as the tent map (Example 1) or the logistic map (Example 2). $\mu_{E}$ and $\mu_{X}$ are typicality measures. They may not be the unique typicality measures, but this does not matter because any other typicality measure agrees on which sets are typical. Note that in my argument typicality is different from probability: the typicality measures do not refer to probabilities in the philosophical sense (they do not refer to, e.g., ontic probabilities describing the distributions which arise in experiments or to degrees of belief).\footnote{Typicality measures are what are formally called ``probability measures'' (see Section~\ref{Typicality} for a formal definition). However, just because they are formally probability measures, this does not mean that they refer to probabilities in the philosophical sense. To give an example: the Lebesgue measure on [0,1] is formally a probability measure, but it is often interpreted as ``length'', which is not probabilistic in the philosophical sense. Similarly, typicality measures which count states do not refer to probabilities. When one says, e.g., that 999/1000 of the balls are red, this statement does not refer to probabilities in the philosophical sense.}\\

\noindent Let me now present a \emph{cost-benefit analysis} of the argument. As outlined when introducing Premise $2$, it can be motivated that the initial probability distributions in SM are translation-continuous. Also, it seems plausible that for any arbitrary open region of phase space one cannot exclude that it might be possible to prepare the system in such a way that there is a positive probability of ending up in this region. Still, these two requirements are not incontestable because the knowledge about these initial probability distributions is limited. Premise 5 (that measures fulfilling Premises 1, 3 and 4 are typicality measures) seems reasonable because all such typicality measures agree on which sets are typical. Still, some might think that these premises are only necessary and not sufficient for typicality measures. Thus they might want to add further requirements (in this case my argument is still relevant: it shows that any typicality measure fulfilling these additional requirements will agree with $\mu_{E}$/$\mu_{X}$ about judgement of typicality).\\

\noindent Concerning the benefits, first, there is a conceptual gain from knowing that the choice of the standard typicality measures can be motivated from Premises $1$-$6$. Second, there is an empirical gain because the typicality measures are related to the initial probability distributions of interest (Premises 3 and 4). In particular, claims about typical behaviour translate into claims about the likely behaviour of the system for any initial probability distribution of interest. Without such a connection, inevitably the question would arise how typicality measures are connected to experiments. Third, according to my argument, a wide range of initial probability distributions are allowed (see Premise 2). This is a strength because, as argued above, different probability distributions may arise in different contexts, and my argument is consistent with this (this is another empirical gain). Particularly the empirical gains might make the argument also attractive for physicists.

\section{A New Proposal: Typicality II}\label{New2}
Let me now outline the analogous argument for the less stringent notion of typicality II.
The first premise remains the same:\vspace{2mm} \\
\noindent\textbf{Premise 1$^{*}$. Typicality measures are invariant under the dynamics.}\\

\noindent Recall that the preparation of a system in SM is described by an initial
probability distribution. Here we assume that these initial probability distributions
$p$ are $\delta*\kappa$-\emph{translation-close}, i.e., that for all open sets $A$ in $X$:
\begin{equation}\label{transclose}
\left|\left|\tau\right|\right|<\delta\rightarrow |p(Tr(A,\tau))-p(A)|<\kappa,
\end{equation}
where $\delta> 0$ and $0<\kappa<1$ are very small real numbers.
The condition of translation-closeness can be motivated in a similar fashion as translation-continuity. That is, when one prepares a system in SM, one does not have sufficient accuracy to create any other probability measure than a translation-close probability measure.\footnote{Vranas (1998) advances another argument for translation-closeness. According to Vranas, measurements correspond to coarse-graining the phase space into finitely many sets (called phase cells) whose states lead to the same measured value. The idea is that in practice all that matters are coarse-grained probability measures (where a measure is assigned only to unions of phase cells) and all that is needed is an argument that any coarse-grained probability measure is translation-close. Vranas's approach does not work because \emph{there are irresolvable technical difficulties}. When the phase space is coarse-grained, then $f_{t}(A)$, where $A$ is a phase cell and $t\in\field{R}^{+}$ or $\field{N}$, will often not be a phase cell again (cf.\ Werndl 2009a). However, the definition of mt-dynamical systems requires that $f_{t}(A)$ is an element of the phase space for all phase cells $A$ and all $t$ (because otherwise the $f_{t}$s are not functions from the phase space to the phase space). Also, because $f_{t}(A)$ and $f^{-1}_{t}(A)$ may not be phase cells again, the conditions of invariance (equations (\ref{I2}) and (\ref{I1})) cease to be applicable. Furthermore, $Tr(A,\tau)$ will usually not be an element of the coarse-grained phase space, implying that the notion of translation-continuity ceases to be applicable. Problems arise even if these technical difficulties are set aside: Vranas claims that as the measurement precision increases, the maximum measure of the size of a phase cell will become smaller. From this he concludes that sufficiently small displacements of sets only lead to a very small change of the coarse-grained measure (implying translation-closeness). However, as van Lith (2001) has pointed out, this conclusion is unwarranted because the decreasing cell size is accompanied by a compensating increase of the number of cells that are added or deleted because of the displacement (hence it is unclear whether the total size of the added or deleted phase cells will become smaller).}
Vranas (1998) has shown a result analogous to Malament and Zabell's equivalence of the conditions of absolute continuity and translation-closeness. To state it, I first need to introduce a definition.
The measure \emph{$\nu$ is $\varepsilon_{1}/\varepsilon_{2}$-continuous w.r.t.\ the measure $\mu$} (where $\nu$ and $\mu$ are both defined on $(X,\Sigma_{X})$) $(\nu_{\varepsilon_{2}}\!\!\ll\!\mu_{\varepsilon_{1}})$ iff:
\begin{equation}
\textnormal{if}\,\,\mu(A) \leq \varepsilon_{1}\,\,\textnormal{for a measurable set}\,\,A, \,\,\textnormal{then}\,\,\nu(A)\leq\varepsilon_{2}.
\end{equation}
The result shown by Vranas is that if $p$ is $\delta*\kappa$-translation-close, $p$ is $\varepsilon_{1}/\varepsilon_{2}$-continuous w.r.t.\ the Lebesgue measure for $\kappa<\varepsilon_{2}<1$ and $\varepsilon_{1}<(\varepsilon_{2}-\kappa)\delta_{n}$, where $\delta_{n}$ is the volume of the $n$-sphere with radius $\delta$.\footnote{Vranas proved this result only for the phase space $\field{R}^{n}$, but it is clear from the proof that it also holds for any arbitrary interval of $\field{R}^{n}$. Although the phase spaces in SM are not intervals, I follow Malament and Zabell (1980) and Vranas (1998) in assuming that this result also holds for systems in SM.}\\

\noindent As before, different initial probability distributions might arise in SM, and thus there is a \emph{class $P$ of initial probability distributions of interest}.
Open sets (and especially open sets whose Lebesgue measure is not very small) are  extended regions of phase space. This motivates the only additional requirement on this class, namely that for an open region which is not very small, one cannot exclude that there might be some way of preparing the system such that the probability of ending up in this region is not very small. Formally, if $\lambda(A)>\Psi$ for an open set $A$ for a very small $\Psi>0$, then $P$ includes an initial probability distribution $p$ with $p(A)>\Psi$. This condition will be required to hold for $\Psi=\varepsilon_{2}$ and $\Psi=\varepsilon_{3}$. So: \vspace{2mm}\\
\noindent \textbf{Premise 2$^{*}$. The initial probability distribution of interest are a class $P$ of translation-close probability distributions where if $\lambda(A)>\Psi$ for an open set $A$, then there is a $p\in P$ with $p(A)>\Psi$ (where $\Psi=\varepsilon_{2}$ or $\varepsilon_{3}$).}\vspace{2mm} \\
Note that \emph{this great flexibility on the initial probability distribution in SM allowed by Premise 2$^{*}$ should be welcome}. \\

\noindent Concerning dynamical systems theory, also in certain applications initial probability distributions over the states play a role. The justification of typicality measures advanced here is relevant whenever Premise 2$^{*}$ holds. For instance, consider again the dynamical system of a frictionless \emph{pendulum}, consisting of a mass $m$ that experiences a single force $F$ which depends only on the position of the mass and a constant. Suppose that one does not have sufficient accuracy to create any other initial probability distribution than a translation-close one (which has some plausibility). Further, since in experiments one has the freedom to prepare the system in all kinds of different initial positions and initial velocities, it seems plausible that for an open region of phase space that is not very small, there is an initial probability distribution which assigns a non-negligible probability to this region. Then the uniform measure can be justified as typicality measure.\\

\noindent
The next two premises \emph{relate typicality to the initial probability distributions of interest}. First, it seems reasonable to require that if a set of states has very small probability for all initial probabilities of interest, then the set is atypical: \vspace{2mm} \\
\noindent \textbf{Premise 3$^{*}$. If $p(A)\leq\varepsilon_{2}$ for all probability distributions of interest for some measurable $A$, then $T(A)\leq\beta$ (for very small $\varepsilon_{2}, \beta$; $0\leq\varepsilon_{2}, \beta$).}\vspace{2mm} \\
\noindent Second, because the typicality measure should serve as a shortcut, it is required that if a set of states is typical, it is of very high probability for all initial probability distributions of interest: \vspace{2mm} \\
\noindent \textbf{Premise 4$^{*}$. If $T(A)> 1-\beta$ for some measurable $A$, then $p(A)> 1-\varepsilon_{3}$ for all probability distributions of interest (for very small $\varepsilon_{3}, \beta$; $0\leq\beta<\varepsilon_{3}$).}\\

\noindent
As before, the following assumption is made:\vspace{2mm} \\
\textbf{Premise 5$^{*}$. Whenever a measure fulfills Premises 1$^{*}$, 3$^{*}$ and 4$^{*}$ (for the probability distributions as characterized by Premise 2$^{*}$), it is a typicality measure}.\vspace{2mm} \\
It will later be shown that measures satisfying Premises 1$^{*}$, 3$^{*}$ and 4$^{*}$ can be interchangeably used as typicality measures.\footnote{This is shown when proving Theorem 2 (see Appendix 2).}  Hence it seems reasonable to take these conditions as sufficient for typicality measures.\\

\noindent Let me specify the systems under consideration:\vspace{2mm} \\
\textbf{Premise 6$^{*}$. Let a statistical mechanical system $(\Gamma_{E},\Sigma_{\Gamma_{E}},\phi_{t},\mu_{E})$ be given, or let a dynamical system $(X,\Sigma_{X},\phi_{t},\mu_{X})$ be given where $X$ is an interval in $\field{R}^{n}$, $\mu_{X \beta}\!\!\ll\!\lambda_{\varepsilon_{2}}$ and $\lambda_{\varepsilon_{4}}\!\!\ll\!\mu_{X \beta}, (\varepsilon_{2}<\beta; \beta<\varepsilon_{4}; \varepsilon_{4}\leq\varepsilon_{3}+\varepsilon_{1}-\varepsilon_{2})$.}\vspace{2mm} \\
For many of the standard measures $\mu_{X}$ in measure-theoretic dynamical systems theory $\mu_{X \beta}\!\!\ll\!\lambda_{\varepsilon_{2}}$ and $\lambda_{\varepsilon_{4}}\!\!\ll\!\mu_{X \beta}$ for very small $\varepsilon_{2},\beta$ and $\varepsilon_{4}$. For instance, this trivially holds for the measure $\mu_{Y}$ of the tent map (Example 1), and it is easy to see that it holds for the measure $\mu_{Z}$ of the logistic map (Example 2).\footnote{Let $\mu_{X}(A)\leq\beta$ for an arbitrary $\beta$. Then $\beta\geq\mu_{X}(A)\geq \int_{A}\omega^{-1} d\lambda=\omega^{-1}\lambda(A)$, where
$\omega^{-1}=\min_{x}\frac{1}{\pi\sqrt{x(1-x)}}$. Hence $\lambda_{\omega\beta}\!\!\ll\!\mu_{X \beta}$. Conversely, let
$\lambda(A)\leq\beta$ for an arbitrary $\beta$. Then $\mu_{X}(A)\leq \int\rho d\lambda\leq \rho\lambda(A)\leq\rho\beta$, where $\rho=\int_{0}^{\beta/2}\frac{1}{\pi\sqrt{x(1-x)}}d\lambda+
\int_{1-\beta/2}^{1}\frac{1}{\pi\sqrt{x(1-x)}}d\lambda$. Thus $\mu_{X\rho\beta}\!\!\ll\!\lambda_{\beta}$.}\\

\noindent Finally, I also adopt the following premise:\vspace{2mm} \\
\textbf{Premise 7$^{*}$. Assume that $\nu_{\beta_{2}}\!\!\ll\!\mu_{\beta_{1}}$ and
$\mu_{\beta_{4}}\!\!\ll\!\nu_{\beta_{3}}$ (for very small $\beta_{i}$, $0<\beta_{i}; 1\leq i\leq 4$) and that $\mu$ and $\nu$ both satisfy Premises 1$^{*}$, 3$^{*}$ and 4$^{*}$. Then, pragmatically speaking, $\mu$ and $\nu$ can be interchangeably used as typicality measures.}\vspace{2mm} \\
This can be motivated as follows. The antecedent of Premise 7$^{*}$ implies that claims about typicality in terms of $\mu$ are translatable into claims about typicality in terms of $\nu$, and vice versa: If a set $D$ is typical with respect to $\mu$, i.e.\ $\mu(D)> 1-\beta_{1}$, $D$ is typical w.r.t.\ $\nu$ because $\nu(D)> 1-\beta_{2}$. Conversely, if a set $D$ is typical with respect to $\nu$, i.e.\ $\nu(D)> 1-\beta_{3}$, $D$ is typical w.r.t.\ $\mu$ because $\mu(D)> 1-\beta_{4}$. That such measures can, pragmatically speaking, be both used as typicality measures seems to be often endorsed by proponents of the typicality approach, e.g., Goldstein (2001) or Maudlin (2007).\\

\noindent
The following result is derivable from these premises:\vspace{2mm} \\
\textbf{
Conclusion (Theorem 2). The measure $\mu_{E}$ of systems in statistical mechanics/the measure $\mu_{X}$ of dynamical systems is a typicality measure. Furthermore, for any other typicality measure $T$, pragmatically speaking, $\mu_{E}$ and $T$/$\mu_{X}$ and $T$ can be interchangeably used as typicality measures.}\vspace{2mm} \\
This result aims to justify typicality measures $\mu_{E}$ in SM and typicality measures $\mu_{X}$ of dynamical systems such as the tent map (Example 1) or the logistic map (Example 2). $\mu_{E}$ and $\mu_{X}$ are typicality measures. They may not be the unique typicality measures, but this does not matter because any other typicality measure can be interchangeably used as typicality measure. Note that, again, the typicality measures do not refer to probabilities in the philosophical sense.\\

\noindent The cost-benefit analysis is similar as for typicality I.
Concerning Premise 2$^{*}$, it seems plausible that the initial probability distributions are translation-close and that for an open region which is not very small, one cannot exclude that there might be some way of preparing the system such that the probability of ending up in this region is not very small. Still, these two requirements are not incontestable because the knowledge about these initial probability distributions is limited. Premise 5$^{*}$ seems reasonable because it follows that all such typicality measures can, pragmatically speaking, be interchangeably used as typicality measures. Still, some might want to add further requirements for something to count as typicality measure (then my argument is still relevant: it shows that measures fulfilling these additional requirements can, pragmatically speaking, be interchangeably used with $\mu_{E}$/$\mu_{X}$ as typicality measures). The motivation of Premise $7^{*}$ is that claims about typicality in terms of $\mu$ are translatable into claims about typicality in terms of $\nu$, and vice versa. Still, $\mu$ and $\nu$ can disagree on the size assigned to sets, and some might not want to allow this.\\

\noindent Regarding the benefits, first, there is a conceptual gain from knowing that the choice of the standard typicality measures follows from Premises $1^{*}$-$7^{*}$. Second, there is an empirical gain because the typicality measures are related to the initial probability distributions of interest (Premises $3^{*}$ and $4^{*}$). Third, it is a strength that a wide range of probability distributions are allowed (see Premise 2$^{*}$), and this is another empirical gain. These empirical gains might make the argument attractive for physicists.

\section{Uniqueness Results}\label{Uniqueness}
\noindent In this section I show that if a further assumption is added to the argument, then the typicality measure is unique (typicality I)/unique from a pragmatic perspective (typicality II). This further assumption is the dynamical condition of ergodicity (typicality I)/epsilon-ergodicity (typicality II). A mt-dynamical system $(X,\Sigma_{X},\mu_{X},f_{t})$ is \emph{ergodic} iff there is no measurable set $A$ in $X$, $0<\mu_{X}(A)<1$, such that $f_{t}(A)=A$ for all $t\in\field{R}^{+}$ or $\field{N}$. A mt-dynamical system $(X,\Sigma_{X},\mu_{X},f_{t})$ is epsilon-ergodic iff it is ergodic on a set of measure $1-\varepsilon_{0}$, for a very small $\varepsilon_{0}\geq 0$.\footnote{Here it is always assumed that $\varepsilon_{0}$ is negligible compared to the measure of any of the macroregions, i.e., that $\varepsilon_{0}/\min_{i}(\mu_{E}(\Gamma_{M_{i}}))=\Theta$, for a very small $\Theta\geq 0$.\label{complication}} Formally: $(X,\Sigma_{X},\mu_{X},f_{t})$ is \emph{$\varepsilon_{0}$-ergodic}, where $0\leq\varepsilon_{0}<1$ is a real number, iff there is a measurable set $V\subset X$, $\mu(V)=1-\varepsilon_{0}$,
with $f_{t}(V)\subseteq V$ for all $t\in\field{R}^{+}$ or $\field{N}$ such that the mt-dynamical system $(V,\Sigma_{V},\mu_{V},f_{t}^{V})$ is ergodic where $\Sigma_{V}=\{V\cap A; A\in\Sigma_{X}\}$, $\mu_{V}(\cdot)=\mu_{X}(\cdot)/\mu_{X}(V)$ for any set in $V$, and $f_{t}^{V}$ is $f_{t}$ restricted to $V$. $(X,\Sigma_{X},\mu_{X},f_{t})$ is \emph{epsilon-ergodic} iff it is $\varepsilon_{0}$-ergodic for a very small $\varepsilon_{0}\geq 0$.\\

\noindent To state the uniqueness results, one more definition is needed. The measures $\mu_{X}$ and $T$ both defined on $(X,\Sigma_{X})$ are \emph{$\theta$-close}, where $0<\theta< 1$ is a very small real number, iff:
\begin{equation}\label{eclose}
|\mu_{X}(A)-T(A)|<\theta\,\,\,\textnormal{for all measurable sets}\,\,A\,\,\,\textnormal{in}\,\,X.
 \end{equation}
Note that for $\theta$-close measures $\mu$ and $\nu$, $\nu_{\beta_{1}+\theta}\ll\mu_{\beta_{1}}$ and $\mu_{\beta_{3}+\theta}\ll\nu_{\beta_{3}}$ for any $\beta_{1},\beta_{3}\geq 0$. Hence, pragmatically speaking, they can be interchangeably used as typicality measures (cf.~Premise~7$^{*}$). Assuming that one does not care about differences of typicality measures of size $\theta$, $\theta$-close measures are \emph{identical from a pragmatic perspective}.
\vspace{2mm} \\
\noindent \textbf{Theorem 3. Suppose that
Premises 1-6 hold and that the statistical mechanical system $(\Gamma_{E},\Sigma_{\Gamma_{E}},\phi_{t},\mu_{E})$/the dynamical system $(X,\Sigma_{X},\phi_{t},\mu_{X})$ is ergodic. Then $\mu_{E}$/$\mu_{X}$ is the unique typicality measure.}
\vspace{2mm} \\
For a proof of Theorem 3, see Appendix 3.
\vspace{2mm} \\
\noindent \textbf{Theorem 4. Suppose that Premises 1$^{*}$-7$^{*}$ hold and that the statistical mechanical system $(\Gamma_{E},\Sigma_{\Gamma_{E}},\phi_{t},\mu_{E})$/the dynamical system $(X,\Sigma_{X},\phi_{t},\mu_{X})$ is $\beta$-ergodic. Then any other typicality measure $T$ is $\varepsilon_{6}$-close to $\mu_{E}$/$\mu_{X}$, where $\varepsilon_{6}=2(\beta+\varepsilon_{4}-\varepsilon_{1})+\beta/(1-\beta)$.}
\vspace{2mm} \\
For a proof of Theorem 4, see Appendix 4. There are many dynamical systems, including the tent map (Example 1) and the logistic map (Example 2) and generally all chaotic systems, which are ergodic/epsilon-ergodic (cf.\ Aligood, Sauer and Yorke 1996; Werndl 2009b). Hence for these systems Theorem 3/Theorem 4 shows that the typicality measure is unique/unique from a pragmatic perspective.\\

\noindent In SM ergodicity and epsilon-ergodicity have a long and notorious history. Boltzmann already invoked the notion of ergodicity in some of his arguments (Uffink 2007). In the contemporary literature on the foundations of SM several papers have defended the claim, or have assumed in their arguments, that systems in SM are ergodic or epsilon-ergodic (e.g., Frigg and Werndl 2012; Malament and Zabell 1980; Pitowsky 2012; Vranas 1998). One of the most important mathematical results so far about ergodicity is the proof of the \emph{Boltzmann-Sinai hypothesis} -- that the motion of $n$ hard-spheres on the two or three-dimensional torus is ergodic (Sim\'{a}nyi 2010). The relevant mathematics is so difficult that for more realistic systems than hard spheres the knowledge is limited and largely based on numerical simulations. Because of this, some argue that one simply does not know yet whether more realistic systems in SM are ergodic or epsilon-ergodic (cf.~Uffink 2007). In this paper no commitment to ergodicity or epsilon-ergodicity is needed. Here it is just important that for systems such as hard spheres Theorem 3/Theorem 4 demonstrates that the typicality measure is unique/unique from a pragmatic perspective. Moreover, if some more realistic systems in SM turn out to be ergodic/epsilon-ergodic (which is certainly possible), Theorem 3/Theorem 4 shows that the typicality measure is unique/unique from a pragmatic perspective.\\

\noindent As a side remark, it should be noted that some have argued that the condition of ergodicity or epsilon-ergodicity guarantees thermodynamic-like behaviour (cf.\ Frigg and Werndl 2011). First, consider ergodicity, which is equivalent to the condition that the portion of time an arbitrary solution stays in $A$ equals the measure of $A$. Formally: $L_{A}(x)=\mu_{E}(A)$
for all initial conditions $x\in\Gamma_{E}$ except, perhaps, a set $B$ with $\mu_{E}(B)=0$, where
\begin{equation}
L_{A}(x)=\lim_{t\rightarrow \infty}\int_{A}\chi_{A}(\phi_{\tau}(x))d\tau
\end{equation}
(here $\chi_{A}(x)$ is the characteristic function\footnote{That is, $\chi_{A}(x)=1$ for $x\in A$ and $0$ otherwise.} of $A$).
Consider an initial condition $x\in\Gamma_{E}\setminus B$. Then the dynamics will carry $x$ to $M_{eq}$ and will keep it there most of the time. The system will move out of the equilibrium region every now and then and visit non-equilibrium regions. Yet since these regions are small compared to the equilibrium region, it will only spend a small fraction of time there. Therefore, the entropy is close to its maximum most of the time and fluctuates away from it only occasionally. Hence if $\mu_{E}$ is interpreted as probability/typicality measure, thermodynamic-like behaviour is of probability $1$/typical (typicality I). Concerning $\varepsilon$-ergodic systems, note that such a system is ergodic on $V$. Consequently, it shows thermodynamic-like behaviour
for the initial conditions in $V$. Then, by the same moves as explained above for ergodicity,
one finds that thermodynamic-like behaviour is of probability $1-\varepsilon$/typical (typicality II).

\section{Conclusion}\label{Conclusion}
A popular view in contemporary Boltzmannian statistical mechanics is to interpret the measures as typicality measures, i.e.\ as representing the relative size of sets of states. In measure-theoretic dynamical systems theory measures can similarly be interpreted as typicality measures. However, a justification why these measures are a good choice of typicality measures is missing, and the paper attempted to contribute to fill this gap.\\

\noindent The paper first criticised Pitowsky (2012) -- the only justification of typicality measures known to the author. Pitowsky's argument goes as follows. Consider the set $\{0,1\}^{\infty}$ of all infinite sequences of zeros and ones. By approximation with the measures defined on the sets $\{0,1\}^{n}$ of finite sequences of zeros and ones, a unique measure $\mu_{\infty}$ is obtained on $\{0,1\}^{\infty}$. Let $\phi$ be the map which assigns to each infinite sequence $\omega$ of zeros and ones the number in the unit interval whose binary development is $\omega$. When $\phi$ is used to map the measure $\mu_{\infty}$ on $\{0,1\}^{\infty}$ to a measure on the unit interval, one obtains the uniform measure. Hence, Pitowsky concludes, the uniform measure is the uniquely correct typicality measure. This paper argued that Pitowsky's argument is problematic. It is unclear why $\phi$ and not another function is used to map the measure $\mu_{\infty}$ to the unit interval. Furthermore, there are counterexamples because for many systems on the unit interval the uniform measure is not the standard measure used and is not invariant.\\

\noindent The paper then provided a first tentative proposal of how to justify typicality measures for two notions, namely typicality I (atypical means measure zero) and typicality II (atypical means very small measure). The main premises of the argument are as follows. The initial probability distributions of interest are translation-continuous (for typicality I) or translation-close (for typicality II). A typicality measure should be related to these initial probability distributions in two ways: First, if a set is of probability zero (for typicality I) or of very small probability (for typicality II) for all probability distributions, then it is atypical. Second, if a set is typical, it is of probability one (for typicality I) or of very high probability (for typicality II)  for all  probability distributions. Furthermore, typicality measures should be invariant. The conclusion are two theorems which show that the standard measures of statistical mechanics and dynamical systems theory are typicality measures. There may be other typicality measures, but these agree on which sets are typical (for typicality I) or can be interchangeably used as typicality measures (for typicality II). Finally, two theorems were presented, showing that if systems are ergodic (for typicality I) or epsilon-ergodic (for typicality II), the typicality measure is unique (for typicality I) or unique from a pragmatic perspective (for typicality II).

\section{Appendix}
\subsection{Proof of Theorem 1}
First of all, consider:
\begin{equation}\label{raus}
\textnormal{if}\,\,\lambda(A)>0,\,\,\textnormal{then there is a }\,\,p\in P\,\,\textnormal{with}\,\,p(A)>0.
\end{equation}
Later in the proof the result is needed that condition~(\ref{raus}) is implied by the condition that if $A$ is an open set, there is a $p\in P$ with $p(A)>0$. This follows because if $\lambda(A)>0$, $A$ contains an open subset $O$ with $\lambda(O)>0$ (since $\lambda$ is regular). Second, note that $(\Gamma_{E},\Sigma_{\Gamma_{E}},\phi_{t},\mu_{E})$ is a dynamical system. $\mu_E$ is defined as:
\begin{equation}\label{MM}
\mu_{E}(A)=\int_{A}\mid\nabla_{x} H\mid^{-1}d\lambda/\int_{\Gamma_{E}}\mid\nabla_{x} H\mid^{-1}d\lambda.
\end{equation}
Hence it follows (from the Radon-Nikodym theorem) that $\mu_{E}\ll\lambda$. Let $\alpha=\min_{x}(\mid\!\!\nabla_{x}H\!\!\mid^{-1})/\int_{\Gamma_{E}}\mid\!\!\nabla_{x} H\!\!\mid^{-1}d\lambda$, $\alpha>0$ (the minimum exists because $\Gamma_{E}$ is compact). Then if $\mu_{E}(A)=0$,  $0=\mu_{E}(A)\geq \alpha \int_{A}d\lambda=\alpha\lambda(A)$, and hence $\lambda(A)=0$. Therefore, also $\lambda\ll\mu_{E}$. Thus it suffices to focus on the case where a dynamical system
$(X,\Sigma_{X},\phi_{t},\mu_{X})$ is given where $\mu_{X}\ll \lambda$ and $\lambda\ll\mu_{X}$. \\

\noindent The definition of a dynamical system requires that $\mu_{X}$ is invariant, and hence $\mu_{X}$ satisfies Premise 1. Next, assume that $p(A)=0$ for all probability distributions of interest for some measurable $A$. Then also $\lambda(A)=0$ (from Premise 2 and condition~(\ref{raus})). Because $\mu_{X}\ll\lambda$ (from Premise 6), $\mu_{X}(A)=0$. Consequently, $\mu_{X}$ satisfies Premise 3. Note that Premise 4 is equivalent to the claim that $p\ll T$ for all probability distributions $p$ of interest. Because $\lambda\ll\mu_{X}$ (from Premise 6), $p\ll\mu_{X}$ for all probability distributions of interest (from Premise 2). Thus $\mu_{X}$ satisfies Premise 4. Because $\mu_{X}$ satisfies Premises 1, 3 and 4, it is a typicality measure (from Premise 5). \\

\noindent Let $T$ be another typicality measure. Then $\lambda\ll T$ (from Premises 2 and 4 and condition~(\ref{raus})). If $\lambda(A)=0$ for some measurable $A$, then $p(A)=0$ for all probability distributions of interest (from Premise 2), and if $p(A)=0$ for all probability distributions of interest, $T(A)=0$ (from Premise 3). Hence $T\ll \lambda$. Therefore, $T(A)=0$ iff $\lambda(A)=0$. Because $\lambda(A)=0$ iff $\mu_{X}(A)=0$ (from Premise 6), $T(A)=0$ iff $\mu_{X}(A)=0$. Hence any other typicality measure $T$ will agree with $\mu_{X}$ about which sets are typical.

\subsection{Proof of Theorem 2}
First of all, consider:
\begin{equation}\label{raus2}
\textnormal{if}\,\,\lambda(A)>\Psi,\,\,\textnormal{then there is a }\,\,p\in P\,\,\textnormal{with}\,\,p(A)>\Psi.
\end{equation}
Later in the proof the result is needed that condition (\ref{raus2}) is implied by the condition that if $\lambda(A)>\Psi$ for an open set $A$, then $P$ includes an initial probability distribution $p$ with $p(A)>\Psi$. This follows because if $\lambda(A)>\Psi$, then there is an open subset $O$ of $A$ with $\lambda(O)>\Psi$ (since $\lambda$ is regular). Second, note that $(\Gamma_{E},\Sigma_{\Gamma_{E}},\phi_{t},\mu_{E})$ is a dynamical system. Vranas (1998) has already shown that $\lambda$ is $\varepsilon/\xi\varepsilon$-continuous w.r.t.\ $\mu_{E}$ for any $\varepsilon\in (0,1)$ and for $\xi=\max_{x}(\mid\!\!\nabla_{x}H\!\!\mid)*\int_{\Gamma_{E}}\mid\!\!\nabla_{x} H\!\!\mid^{-1}d\lambda$ (the maximum exists because $\Gamma_{E}$ is compact).
Let $\lambda(A)\leq\varepsilon$, $\varepsilon\in (0,1)$. Then from equation~(\ref{MM}) it follows that $\mu_{E}(A)\leq\chi\int_{A}d\lambda=\chi\lambda(A)\leq\chi\varepsilon$ for $\chi\!\!=\!\!\max_{x}(\mid\!\!\nabla_{x}H\!\!\mid^{-1})/\int_{\Gamma_{E}}\mid\!\!\nabla_{x} H\!\!\mid^{-1}d\lambda$ (the maximum exists because $\Gamma_{E}$ is compact). Hence $\mu_{E}$ is $\varepsilon/\chi\varepsilon$-continuous w.r.t.\ $\lambda$. Thus it suffices to focus on the case where a dynamical system $(X,\Sigma_{X},\phi_{t},\mu_{X})$ is given with $\mu_{X\beta}\!\!\ll\!\lambda_{\varepsilon_{1}}$ and $\lambda_{\varepsilon_{4}}\!\!\ll\!\mu_{X \beta}$.\\

\noindent Premise 1$^{*}$ holds because for a dynamical system $\mu_{X}$ is invariant. Next, assume that $p(A)\leq\varepsilon_{2}$ for all probability distributions of interest for some measurable $A$. Then $\lambda(A)\leq\varepsilon_{2}$ (from Premises 2$^{*}$ and condition~(\ref{raus2})). Hence $\mu_{X}(A)\leq\beta$ (from Premise 6$^{*}$), and $\mu_{X}$ satisfies Premise 3$^{*}$. Premise 4$^{*}$ is equivalent to the claim that $p_{e_{3}}\ll T_{\beta}$ for all probability distribution of interest.
Suppose that $\mu_{X}(A)\leq\beta$. Then $\lambda(A)\leq\varepsilon_{4}$ (from Premise 6$^{*}$) and $p(A)\leq\varepsilon_{2}+(\varepsilon_{4}-\varepsilon_{1})\leq\varepsilon_{3}$ (from Premise 2$^{*}$). Therefore, $\mu_{X}$ satisfies Premise 4$^{*}$. Because $\mu_{X}$ satisfies Premises 1$^{*}$, 3$^{*}$ and 4$^{*}$, it is a typicality measure (from Premise 5$^{*}$.\\

\noindent Let $T$ be another typicality measure. If $T(A)\leq\beta$, then
$\lambda(A)\leq\varepsilon_{3}$ (from Premise 2$^{*}$ and 4$^{*}$ and equation~(\ref{raus2})). Thus, $\mu_{X}(A)\leq\beta+(\varepsilon_{3}-\varepsilon_{2})$ (from Premise 6$^{*}$). Conversely, if $\mu_{X}(A)\leq\beta$, $\lambda(A)\leq\varepsilon_{4}$ (from Premise 6$^{*}$). Hence $p(A)\leq\varepsilon_{2}+(\varepsilon_{4}-\varepsilon_{1})$ for all probability distributions of interest (from Premise 2$^{*}$), and
$T(A)\leq\beta+(\varepsilon_{4}-\varepsilon_{1})$ (from Premise 3$^{*}$).
Consequently, pragmatically speaking, $\mu_{X}$ and $T$ can be interchangeably used as typicality measures.

\subsection{Proof of Theorem 3}
As for the reasons given when proofing Theorem 1, it suffices to focus on the case where a dynamical system $(X,\Sigma_{X},\phi_{t},\mu_{X})$ is given where $\mu_{X}\ll \lambda$ and $\lambda\ll\mu_{X}$. According to a theorem in ergodic theory, if $(X,\Sigma_{X},\phi_{t},\mu_{X})$  is ergodic and $T$ is a measure invariant under the dynamics with $T\ll \mu_{X}$, then $T=\mu_{X}$ (cf.\ Cornfeld et al.\ 1982). Any other typicality measure $T$ is invariant (from Premise 1). Also, $T\ll \lambda$  (from Premises 2 and 3), $\lambda\ll\mu_{X}$ (from Premise 6), and hence $T\ll\mu_{X}$. Consequently, $T=\mu_{X}$.

\subsection{Proof of Theorem 4}
As for the reasons given when proofing Theorem 2, it suffices to focus on the case where a dynamical system $(X,\Sigma_{X},\phi_{t},\mu_{X})$ is given where $\mu_{X\beta}\!\!\ll\!\lambda_{\varepsilon_{1}}$ and $\lambda_{\varepsilon_{4}}\!\!\ll\!\mu_{X \beta}$. According to a theorem by Vranas (1998), if $(X,\Sigma_{X},\phi_{t},\mu_{X})$ is $\beta$-ergodic, $T$ is invariant and  $T_{\beta+\varepsilon_{4}-\varepsilon_{1}}\ll\mu_{X\beta}$, then $\mu_{X}$ and $T$ are $\varepsilon_{6}$-close with  $\theta=2(\beta+\varepsilon_{4}-\varepsilon_{1})+\beta/(1-\beta)$.
Let $T$ be any other typicality measure. Then $T$ is invariant (from Premise $1^{*}$). Also, $\lambda_{\varepsilon_{4}}\ll \mu_{X \beta}$ (from Premise $6^{*}$),
$T_{\beta+(\varepsilon_{4}-\varepsilon_{1})}\ll\lambda_{\varepsilon_{4}}$ (from Premises 2$^{*}$ and 3$^{*}$), and hence $T_{\beta+(\varepsilon_{4}-\varepsilon_{1})}\ll\mu_{X \beta}$. Consequently, $\mu_{X}$ and $T$ are $\theta$-close.

\section*{References}
\addcontentsline{toc}{section}{References}
\begin{list}{}{    \setlength{\labelwidth}{0pt}
    \setlength{\labelsep}{0pt}
    \setlength{\leftmargin}{24pt}
    \setlength{\itemindent}{-24pt}
  }

\item Aligood, K., Sauer, T. and Yorke, J. (1996), {\em Chaos: An Introduction to Dynamical Systems}, New York: Springer.

\item Cornfeld, I.\ P., Fomin, S.\ V.\ and Sinai, Ya.\ G.\ (1982), \emph{Ergodic Theory}, Berlin et al.: Springer.

\item Davey, K. (2008), `The justification of
  probability measures in statistical mechanics', {\em Philosophy of Science}
  {\bf 75},~28--44.

\item D\"{u}rr, D. (1998), `\"{U}ber den {Zufall}
  in der {Physik}', paper given at the 1998 {Leopoldina Meeting, Halle}.\\
  http://www.mathematik.uni-muenchen.de/~duerr/zufall/zufall.html.

\item Frigg, R. (2008), A field guide to recent
  work on the foundations of statistical mechanics, {\em in} D.~Rickles, ed.,
  `The Ashgate Companion to Contemporary Philosophy of Physics', Ashgate,
  London, pp.~99--196.

\item Frigg, R. and Werndl, C. (2011),
  `Explaining thermodynamic-like behaviour in terms of epsilon-ergodicity',
  {\em Philosophy of Science} {\bf 78},~628--652.

\item Frigg, R. and Werndl, C. (2012), `Demystifying typicality', \emph{Philosophy of Science} \textbf{5}, 917--929.

\item Goldstein, S. (2001), {Boltzmann's} approach
  to statistical mechanics, {\em in} J.~Bricmont, D.~D\"{u}rr, M.~Galavotti,
  G.~Ghirardi, F.~Pettrucione and N.~Zanghi, eds, `Chance in Physics:
  Foundations and Perspectives', Springer, Berlin and New York, pp.~39--54.

\item Goldstein, S. and Lebowitz, J. (2004), `On the ({Boltzmann}) entropy of nonequilibrium systems', {\em Physica D}
  {\bf 193},~53--66.

\item Hemmo, M. and Shenker, O. (2012),
  Measures over initial conditions, {\em in} J.~Ben-Menahem and
  M.~Hemmo, eds, `Probability in Physics', The Frontiers Collection, Springer,
  Berlin and New York, pp.~87--98.

\item Lavis, D. (2005), `Boltzmann and {Gibbs}: An
  attempted reconciliation', {\em Studies in History and Philosophy of Modern
  Physics} {\bf 36},~245--273.

\item Lebowitz, J.~L. (1993a), `Boltzmann's
  entropy and time's arrow', {\em Physics Today} {\bf September Issue},~32--38.

\item Lebowitz, J.~L. (1993b), `Macroscopic
  laws, microscopic dynamics, time's arrow and {Boltzmann's} entropy', {\em
  Physica A} {\bf 194},~1--27.

\item Lebowitz, J.~L. (1999), `Statistical
  mechanics: A selective review of two central issues', {\em Reviews of Modern
  Physics} {\bf 71},~346--357.

\item Leeds, S. (1989), `Malament and {Zabell} on
  {Gibbs} phase averaging', {\em Philosophy of Science} {\bf 56},~325--340.

\item Malament, D. and Zabell, S. (1980),
  `Why {Gibbs} phase averages work - the role of ergodic theory', {\em
  Philosophy of Science} {\bf 47},~339--349.

\item Maudlin, T. (2007), `What could be objective
  about probabilities?', {\em Studies in History and Philosophy of Modern
  Physics} {\bf 38},~275--291.

\item Nielsen, O.\ A.\ (1997), \emph{An Introduction to Integration and Measure Theory}, Wiley-Interscience, New York.

\item Petersen, K. (1983), {\em Ergodic Theory},
  Cambridge University Press, Cambridge.

\item Pitowsky, I. (2012), Typicality and the role
  of the {Lebesgue} measure in statistical mechanics, {\em in} J.~Ben-Menahem
  and M.~Hemmo, eds, `Probability in Physics', The Frontiers
  Collection, Springer, Berlin and New York, pp.~51--58.

\item Sim\'{a}nyi, N. (2010), `The Boltzmann--Sinai Ergodic Hypothesis In Full Generality', arXiv:1007.1206v2 [math.DS].

\item Uffink, J. (1996), `Nought but Molecules in Motion, a review essay of L. Sklar's ``Physics and Chance''', \emph{Studies in History and Philosophy of Modern Physics} \textbf{27}, 373--378.

\item Uffink, J. (2007), Compendium to the
  foundations of classical statistical physics, {\em in} J.~Butterfield
  and J.~Earman, eds, `Philosophy of Physics', North-Holland.,
  Amsterdam, pp.~923--1074.

\item van Lith, J. (2001), Stir in Stilness, PhD
  thesis, University of Utrecht.

\item Volchan S.~B. (2007), `Probability as
  typicality', {\em Studies in History and Philosophy of Modern Physics} {\bf
  38},~801--814.

\item Vranas, P.~B. (1998), `Epsilon-ergodicity and
  the success of equilibrium statistical mechanics', {\em Philosophy of
  Science} {\bf 65},~688--708.

\item Werndl, C. (2009a), `Are deterministic
  descriptions and indeterministic descriptions observationally equivalent?',
  {\em Studies in History and Philosophy of Modern Physics} {\bf 40},~232--242.

\item Werndl (2009b), `What are the new
  implications of chaos for unpredictability?', {\em The British Journal for
  the Philosophy of Science} {\bf 60},~195--220.

\end{list}
\end{document}